\documentclass[11pt,twoside]{article}
\usepackage{asp2014}

\aspSuppressVolSlug
\resetcounters

\begin{document}

\title{Making organizational software easier to find in ASCL and ADS}

\author{Alice~Allen,$^{1,2}$ Siddha Mavuram,$^2$ Robert J. Nemiroff,$^3$  Judy Schmidt,$^1$ and Peter Teuben$^2$}
\affil{$^1$Astrophysics Source Code Library; \email{aallen@ascl.net}}
\affil{$^2$University of Maryland, College Park, MD, USA}
\affil{$^3$Michigan Technological University, Houghton, MI, USA}

\paperauthor{Alice~Allen}{aallen@ascl.net}{0000-0003-3477-2845}{Astrophysics Source Code Library/University of Maryland College Park}{Astronomy Department}{College Park}{MD}{20742}{USA}
\paperauthor{Siddha~Mavuram}{}{}{}{University of Maryland}{College Park}{MD}{20742}{USA}
\paperauthor{Robert~J.~Nemiroff}{nemiroff@mtu.edu}{}{}{Michigan Technological University}{Houghton}{MI}
{}{USA}
\paperauthor{Judy~Schmidt}{gecko@geckzilla.com}{}{Astrophysics Source Code Library}{}{}{}{}{}
\paperauthor{Peter~Teuben}{teuben@astro.umd.edu}{0000-0003-1774-3436}{Astronomy Department}{University of Maryland}{College Park}{MD}
{20742}{USA}




%
  
\begin{abstract}

Software is the most used instrument in astronomy, and organizations such as NASA and the Heidelberg Institute for Theoretical Physics (HITS) fund, develop, and release research software. NASA, for example, has created sites such as code.nasa.gov to share its software with the world, but how easy is it to see what NASA has? Until recently, searching NASA's Astrophysics Data System (ADS) for NASA astronomy research software has not been fruitful. Through its ADAP program, NASA funded the Astrophysics Source Code Library to improve the discoverability of these codes. Adding institutional tags to ASCL entries makes it easy to find this software not only in the ASCL but also in ADS and other services that index the ASCL. This presentation covered the changes the ASCL made as a result of this funding and how you can use the results of this work to better find organizational software in ASCL and ADS.
  
\end{abstract}

\section{Introduction}

\begin{figure}[ht]
  \begin{center}
  \includegraphics[width=.8\linewidth]{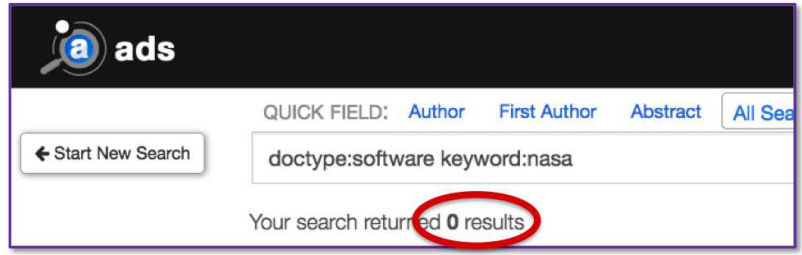}
  \end{center}
  \caption{Screenshot of results for a 2017 ADS search for NASA software}
  \label{fig:noresults}
\end{figure}

In 2017, the editor of the Astrophysics Source Code Library (ASCL),\footnote{\url{https://ascl.net}} Alice Allen, was searching NASA's Astrophysics Data System (ADS)\footnote{\url{https://ui.adsabs.harvard.edu/}} to see what she could learn about software and software use in astrophysics research. As a result of a 2015 ADASS BoF session, in which a suggestion was made for ADS to include software in its categorization of entries \citep{2017ASPC..512..675A}, ADS had added a document type value for software \citep{2019ASPC..521..737T}, so she was experimenting with that, and did a search for ``doctype:software keyword:NASA.'' There were zero hits (Figure \ref{fig:noresults}). This was perhaps not surprising, since ``doctype:software'' was relatively new, but made Allen experiment more, using NASA software as the object of her search. This further experimentation made it clear that it was difficult to search for and find NASA software in ADS. Exacerbating the situation is having to search multiple NASA sites, such as \url{https://heasarc.gsfc.nasa.gov/docs/software.html} and  \url{https://software.nasa.gov}, to discover NASA research software. Thus, a project idea arose: make it easier to find software developed at or paid for by a funding organization in the ASCL by using keywords and pass this information to ADS, which already ingests the ASCL and classifies its entries as the ``software'' document type. 

During a NASA Astrophysics Data Analysis Program (ADAP) funding call, we proposed looking for NASA astrophysics research software on NASA sites such as \url{code.nasa.gov} and \url{software.nasa.gov}, adding these codes to the ASCL if an entry for the software did not already exist, and tagging the new or existing ASCL entries with the keyword ``NASA'' and, when appropriate, with the names of the mission for which the software was written or used. Our proposal was accepted; this poster shared the work we've done and our results so far.
\section{Changes to the ASCL}
The ASCL is very careful about the quantity of metadata it stores, as the more metadata is included in the resource, the more maintenance is needed to keep the records up-to-date \citep{2015JORS....3E..15A}; as a result, the Library did not have a keyword field. The NASA ADAP project required that the ASCL be modified to add this field. Not only did this require changes to our database structure, it also required changes to our input and editing forms and code entry display screens. Further, we had to ensure that information we stored in the keyword field would flow to, and be used by, ADS, Web of Science, and other indexers that ingest our records. The designer and developer of the ASCL, Judy Schmidt, was responsible for making these changes to the ASCL infrastructure; she also created reporting mechanisms to easily view the ``NASA'' tagged software on the ASCL. Siddha Mavuram, a University of Maryland computer science major with an interest in astronomy, was hired to develop software and tracking methods for mining NASA software sites. He also developed an API for the ASCL \citep{P9-103_adassxxx} to provide a way to gather the ASCL's information for NASA software programmatically. Peter Teuben provided additional programming and directed Mavuram's activities, and Robert Nemiroff, the founder of the ASCL, provided guidance and served as a sounding board for infrastructure changes.

\section{Tagging existing ASCL records, searching NASA code sites}
Searching for suitable software took place in two phases; first, Allen looked at existing ASCL entries to see which were hosted on NASA websites or indicated funding from NASA and tagged those entries with ``NASA'' and, if appropriate, mission keywords. She then started mining NASA code sites for research software that meets the ASCL's criteria,\footnote{https://ascl.net/wordpress/submissions/editiorial-policy/} aided by tools developed by Teuben and Mavuram; this work is ongoing and will continue through the end of the project.

\section{Results}
\begin{figure}[ht]
  \begin{center}
  \includegraphics[width=0.8\linewidth]{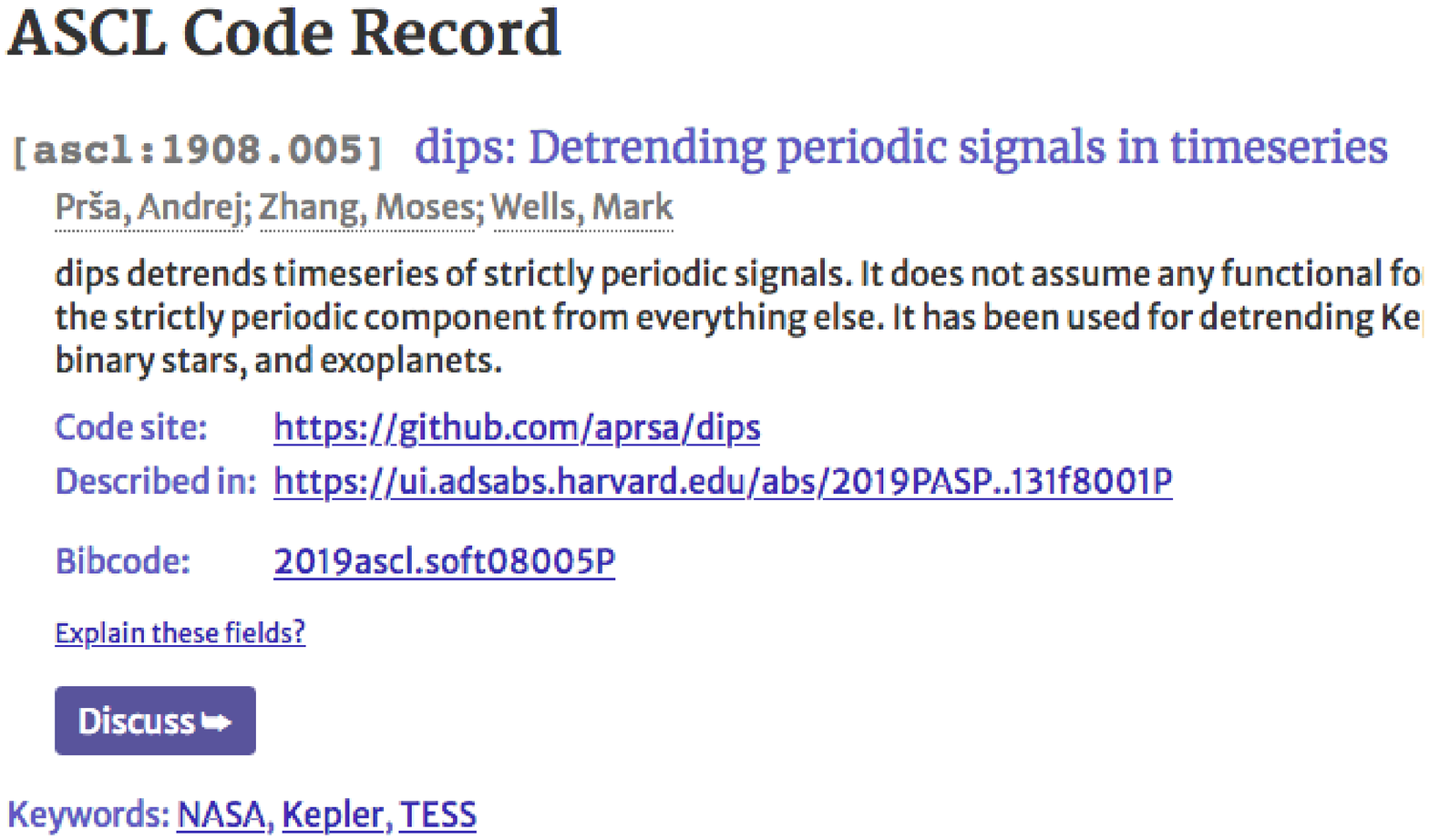}
  \end{center}
  \caption{ASCL record showing keywords for NASA and NASA missions}
  \label{fig:dipsASCLrecordscreenshot}
\end{figure}

Keywords to associate NASA and NASA missions with software now appear on ASCL (Figure \ref{fig:dipsASCLrecordscreenshot}) and ADS records. 

We can find NASA astronomy research software on the ASCL with a keyword search. Because ADS picks up ASCL keywords when it ingests the ASCL's records, it is also possible to search ADS for this software by using ``doctype:software keyword:NASA'' as the search terms. ASCL entries are citable and ADS tracks citations to ASCL entries. 
\begin{figure}[h!]
  \begin{center}
  \includegraphics[width=0.8\linewidth]{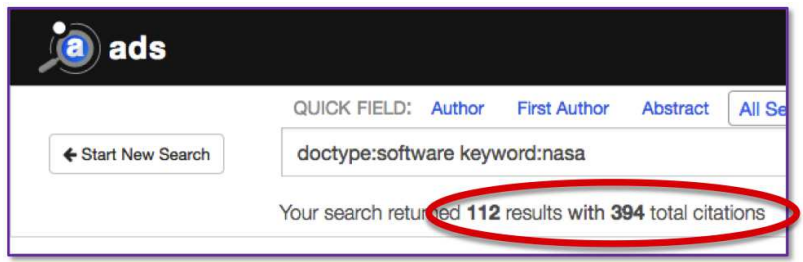}
  \end{center}
  \caption{Screenshot of a 2020 ADS search results for NASA software}
  \label{fig:results}
\end{figure}
As a result, NASA can see what research has been enabled by making its software public. Through the number of citations (Figure \ref{fig:results}), NASA also has a measure (albeit an incomplete one) of the impact funding software development can have on scientific discovery.

\begin{figure}[ht]
\begin{center}
  \includegraphics[width=0.5\linewidth]{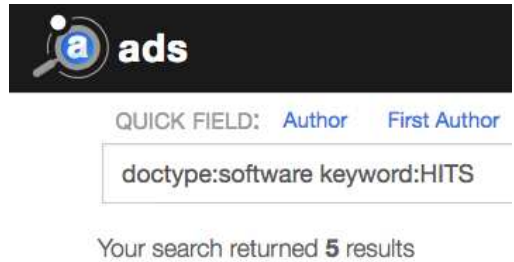}
  \end{center}
  \caption{Screenshot of a 2020 ADS search results for HITS software}
  \label{fig:HITS}
\end{figure}

The changes we have made to the ASCL as a result of this project also provides an opportunity to improve discovery of other institutional software, such as that funded by the Heidelberg Institute for Theoretical Studies, as is shown by Figure \ref{fig:HITS}. 

\section{Conclusion}
Discovery of organization-funded or -written software can be improved by leveraging the ASCL and ADS through the use of keywords. This discoverability can improve use of the software, and also provides information as to the impact the software may have on scientific discovery by linking the software to research which used the software.

\acknowledgements This project was funded by NASA award NNH17ZDA001N-ADAP. ASCL is supported by Michigan Technological University, University of Maryland College Park, and Heidelberg Institute for Theoretical Studies.

\bibliography{P10-212}


\end{document}